\documentclass[aps,pra,twocolumn,showpacs,amsmath,amsmath,amssymb]{revtex4-1}
\usepackage{amsmath,amssymb,graphicx,bm,psfrag,bbold,float,color}
\usepackage{graphicx,physics,hyperref,appendix}
\usepackage{braket}
\usepackage[mathscr]{eucal}

\DeclareMathAlphabet{\mathpzc}{OT1}{pzc}{m}{it}

\begin{document}

\title{Open Ising Model Perturbed by Classical Colored Noises}

\author{Deshui Yu$^{1}$ and Rainer Dumke$^{1,2}$}
\address{\mbox{$^{1}$Centre for Quantum Technologies, National University of Singapore, 3 Science Drive 2, Singapore 117543, Singapore}}
\address{$^{2}$Division of Physics and Applied Physics, Nanyang Technological University, 21 Nanyang Link, Singapore 637371, Singapore}

\begin{abstract}
We investigate the non-Markovian dynamics of an open Ising model simulated by a superconducting circuit. The quantum many-body system is weakly coupled to a white, pink- or blue-colored environment. The relaxation of the system in the strong inter-qubit interaction regime shows a metastable behaviour. In comparison with the dissipative system in the Markovian limit, the negative memory of the blue-colored noise weakens the system's relaxation. However, for the pink-colored noise the relaxation rate of the system is enhanced due to the positive memory effect. The understanding of quantum many-body systems responding to different colored noise fields is necessary for designing the environment of superconducting qubits in a large scale quantum processor.
\end{abstract}


\maketitle

\section{Introduction}~\label{Introduction}


Simulating a quantum system via another mathematically identical system possesses wide applications in condensed-matter physics, high-energy physics and quantum chemistry~\cite{RMP:Georgescu2014}. The successful simulation platforms range from ultracold atoms~\cite{NatPhys:Bloch2012}, trapped ions~\cite{NatPhys:Blatt2012}, photonic systems~\cite{NatPhys:AspuruGuzik2012}, to superconducting (SC) circuits~\cite{NatPhys:Houck2012}. In the past, large resources have been devoted to maximally isolate quantum physical systems from the environment (reservoir). However, the interaction with the environment still dictates the evolution of quantum systems. Recently, an increasing interest has been attracted on the study of many-body open quantum systems for the applications in quantum computing~\cite{NatComm:Reiter2017}, exotic state preparation~\cite{NatPhys:Verstraete2009}, driven-dissipation phase transition~\cite{PRL:Lee2012} and quantum memory~\cite{PRA:Pastawski2011}. In particular, more and more attention is being paid to the theory of non-Markovian processes~\cite{PRA:Jing2013,PRA:Overbeck2016,PRA:Sweke2016,PRA:Mascarenhas2017,PRL:Chenu2017,NewJPhys:Banchi2018}, where quantum systems can receive the information and energy back from the environment, i.e., the memory effect.

Two common approaches have been developed to treat this memory effect. Firstly, the memory kernel master equations based on phenomenologically introducing a time-convolution dynamic term $\int_{0}^{t}ds{\cal{K}}(t,s)\rho(s)$, where $\rho(t)$ is the system's density matrix operator and the two-time operator ${\cal{K}}(t,s)$ acts on the system's Hilbert space~\cite{Nakajima1958,JChemPhys:Zwanzig1960,PRA:Shabani2005}. Secondly, local-in-time master equations derived from the time-convolutionless projection operator method~\cite{ZPhysB:Chaturvedi1979,JStatPhys:Shibata1977}. The dynamic term in master equations is written in the form $\sum_{i}\frac{\gamma_{i}(t)}{2}[2A_{i}(t)\rho A^{\dag}_{i}(t)-\{A^{\dag}_{i}(t)A_{i}(t),\rho\}]$. Here the time-dependent decay rates $\gamma_{i}(t)$ can have temporarily negative values, which actually encode the system-evolution history. The practical usefulness of the former approach is limited due to the difficulty in the evaluation of the memory integral, and also the memory kernel alone does not guarantee the non-Markovian character~\cite{PRA:Mazzola2010}. In contrast, the latter approach is mathematically simple and has been widely applied to various non-Markovian problems~\cite{PRL:Piilo2008,PRB:Ferraro2008,PRA:Piilo2009,JChemPhys:Rebentrosta2009}.

Studying dissipative dynamics of quantum many-body systems in the non-Markovian limit is of much interest currently. Conventional methods of engineering many-body Hamiltonians~\cite{PRL:Ajoy2013,NJP:Hayes2014,SciRep:DiCandia2015} and the system-environment interaction (i.e., quantum jump operators) have alreadly provided various avenues for understanding many-body open quantum systems described by Lindblad-type master equations~\cite{NatPhys:Diehl2008}. Direct tailoring the environment offers an extra opportunity to explore the non-equilibrium behavior of quantum systems beyond the Markov limit~\cite{PRA:Lloyd2001,NatPhys:Verstraete2009,NatCommun:Boixo2013}. Nonetheless, it is still far from clear how a noise in specific color affects the dissipative dynamics of quantum many-body systems. The main reason lies in the absence of effective measures to engineer the environment. However, the situation is different for SC quantum circuits influenced by classical colored noise fields~\cite{PRL:Chenu2017}.

Owing the unique features, like flexibility, tunability and scalability~\cite{Nature:Nakamura1999,Science:Chiorescu2003,PRA:Koch2007,RPP:Wendin2017,JPC:Yu2018,PRA:Yu2018_2}, SC quantum circuits based on Josephson junctions (JJs) enable to simultaneously engineer the Hamiltonian, the environment and their interaction of a many-body system. This makes these solid-state devices also suitable for use as the platforms simulating dissipative many-body quantum systems. The relevant physical parameters may be tuned from the weak-coupling to the ultrastrong-coupling regime~\cite{PRA:Yu2016_1,PRA:Yu2016_2,SciRep:Yu2016,PRA:Yu2017,QST:Yu2017,NJP:Yu2018,PRA:Yu2018,NatPhys:Yoshihara2017}, where the traditional quantum gaseous and photonic platforms have never accessed before. Meanwhile, the noise spectrum of the environment can be altered to a specific color via conventional signal processing techniques~\cite{RJAV:Zhivomirov2018}. In addition, hybridizing solid-state devices and neutral or charged particles bridges the information communication between macroscopic and microscopic quantum systems.

\begin{figure}
\includegraphics[width=8.5cm]{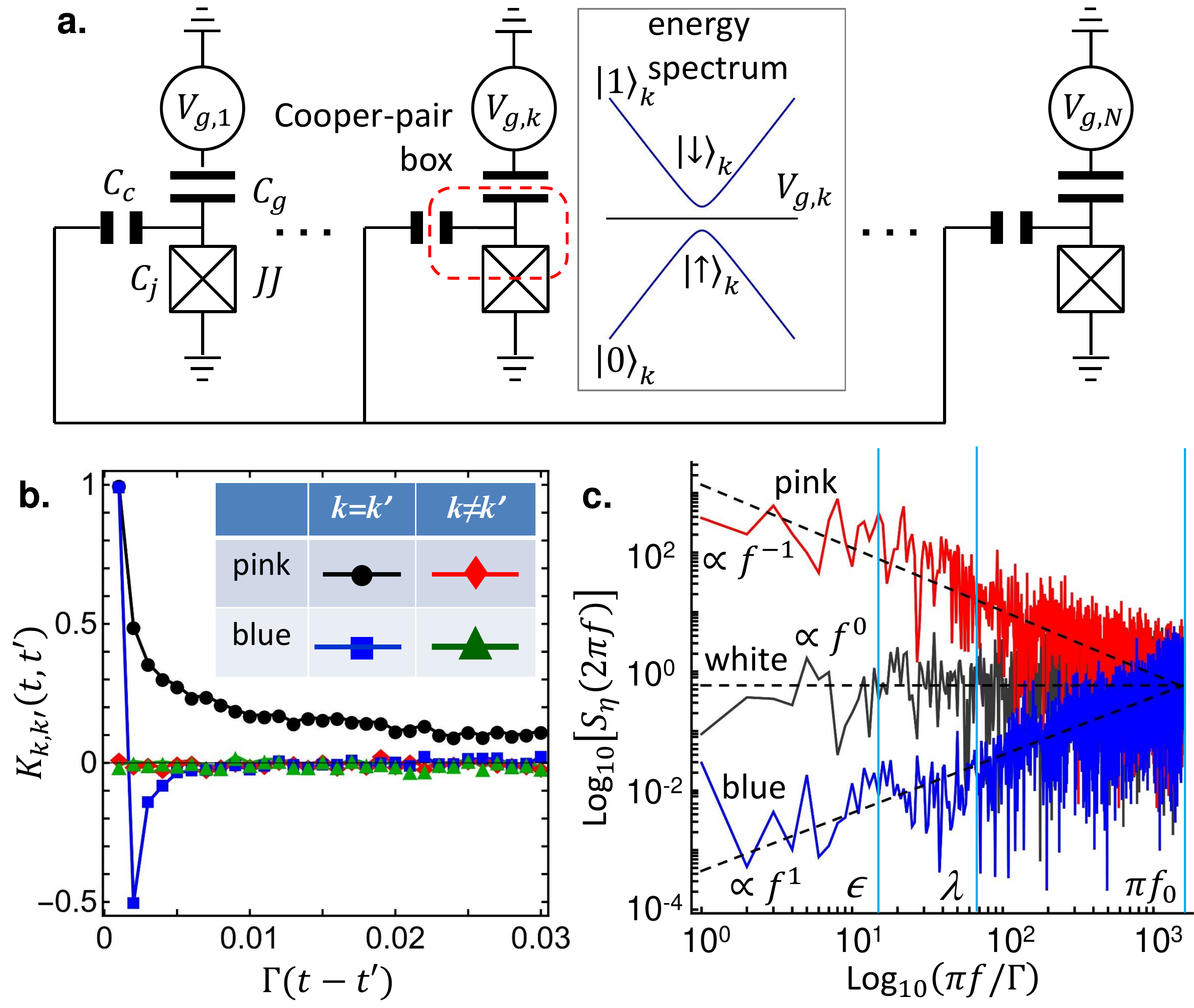}\\
\caption{(Color online) SC-circuit-based Ising model. (a) Multi-charge-qubit system. Each qubit consists of a Cooper-pair box coupled to the Cooper-pair reservoir via a JJ. All qubits have same gate capacitance $C_{g}$ and same self-capacitance of JJs $C_{j}$ and interact with each other via identical coupling capacitors $C_{c}$. A voltage source $V_{g,k}$ is applied to tune the energy spectrum of the $k$-th box. At $\frac{C_{g}V_{g,k}}{2e}=N_{0}+\frac{1}{2}$ with an integer $N_{0}$, two qubit states $\Ket{\uparrow}_{k}=\frac{1}{\sqrt{2}}(\Ket{1}_{k}+\Ket{0}_{k})$ and $\Ket{\downarrow}_{k}=\frac{1}{\sqrt{2}}(\Ket{1}_{k}-\Ket{0}_{k})$ form a spin. $V_{g,k}$ is perturbed by an external noise $\delta V_{g,k}(t)$ around a central value $V_{0}$. Stochastic field $\eta_{k}(t)$, which is proportional to $\delta V_{g,k}(t)$, can be tailored into different colored noise fields. (b) Examples of correlation function $K_{k,k'}(t,t')$ vs. time difference $(t-t')$ for pink- and blue-colored noise fields. $K_{k,k'}(t,t')$ is in units of $f_{0}$. (c) Examples of PSD of white, pink- and blue-colored noise fields $S_{\eta}(2\pi f)$ as a function of noise frequency $f$ (solid lines). Dashed lines indicate the expected frequency dependencies, i.e., $\propto f^{0}$, $\propto f^{-1}$ and $\propto f^{1}$ for white, pink- and blue-colored noise fields, respectively.}\label{Fig1}
\end{figure}

Here we study an experimentally-feasible scheme for the simulation of a quantum Ising model coupled to an artificially-tailored environment. An ensemble of fully-coupled SC qubits is perturbed by classical white, pink- or blue-colored noise. The dissipative dynamics of open quantum system is investigated by numerically solving the master equation which has a local-in-time form. The results indicate that in comparison with the memoryless white-noise perturbation, the relaxation of the time evolution of collective spin polarization is accelerated in the pink-colored environment because of the positive memory, while the negative memory effect slows down the relaxation of the system in the blue-colored environment. 

\section{Physical Model}\label{Physical_Model}

The quantum-simulation platform for the Ising system is composed of $N$ identical single-JJ charge qubits (Cooper-pair boxes)~\cite{Nat:Nakamura1999} (see Fig.~\ref{Fig1}a). The $k$-th qubit is biased by a voltage source $V_{g,k}$ via a gate capacitor $C_{g}$. Identical capacitors $C_{c}$ are used to link all boxes. The charging energy for Cooper pairs in the box is given by $E_{C}=\frac{(2e)^{2}}{2C_{\Sigma}}$ with the total capacitance $C_{\Sigma}=C_{g}+C_{j}+C_{c}$, where $C_{j}$ is the self-capacitance of JJ. The Josephson energy of Cooper pairs is $E_{J}$ ($\ll E_{C}$). The voltage bias $V_{g,k}$ is artificially perturbed by an external weak noise,
\begin{equation}
V_{g,k}(t)=V_{0}+\delta V_{g,k}(t).
\end{equation}
The central value $V_{0}$ is set at $\frac{C_{g}V_{0}}{2e}=N_{0}+\frac{1}{2}$ with an integer $N_{0}$ while the colored noise $\delta V_{g,k}(t)$ may be tailored by filtering a white noise signal with a Nyquist frequency $\pi f_{0}$~\cite{RJAV:Zhivomirov2018}. We assume that all $\{\delta V_{g,k}(t);k=1,...,N\}$, although independent of each other, are in the same color and have the same amplitude.

In the two-state approximation, where $\Ket{0}_{k}$ and $\Ket{1}_{k}$ represent the absence and presence of a single excess Cooper pair in the $k$-th box, the Hamiltonian of many-body system is derived as (see Appendix~\ref{App_A})
\begin{equation}\label{Hamiltonian1}
H=H_{s}+\hbar\sqrt{\Gamma/2}\sum_{k}\eta_{k}(t)\sigma^{x}_{k}.
\end{equation}
The system's coherent dynamics is governed by
\begin{equation}
H_{s}/\hbar=-\epsilon\sum_{k}\sigma^{z}_{k}+(\lambda/N)\sum_{k<k'}\sigma^{x}_{k}\sigma^{x}_{k'}.
\end{equation}
The $x$- and $z$-components of the Pauli operator for the $k$-th qubit are given by $\sigma^{x}_{k}=(\Ket{\uparrow}\Bra{\downarrow})_{k}+(\Ket{\downarrow}\Bra{\uparrow})_{k}$ and $\sigma^{z}_{k}=(\Ket{\uparrow}\Bra{\uparrow})_{k}-(\Ket{\downarrow}\Bra{\downarrow})_{k}$ with two spin states $\Ket{\uparrow}_{k}=\frac{1}{\sqrt{2}}(\Ket{1}_{k}+\Ket{0}_{k})$ and $\Ket{\downarrow}_{k}=\frac{1}{\sqrt{2}}(\Ket{1}_{k}-\Ket{0}_{k})$. The term $H_{s}$ is equivalent to the quantum Ising model~\cite{BOOK:Sachdev1999}, except that each qubit interacts with others equally, i.e., the fully-coupled many-body system. The term $\epsilon=\frac{E_{J}}{2\hbar}$ plays the role of the external magnetic field and gives the spin-state separation in energy. The term $\lambda=\frac{E_{C}}{2\hbar}\frac{C_{c}}{C_{g}+C_{j}}$ corresponds to the interqubit coupling strength and can be enhanced by increasing the coupling capacitance $C_{c}$. As a result, $\lambda$ is tunable in the range from the weak- ($\lambda/\epsilon\ll1$) to the strong-coupling ($\lambda/\epsilon\gg1$) limit.

The real stochastic fields $\{\eta_{k}(t);k=1,...,N\}$ in Hamiltonian~(\ref{Hamiltonian1}) depend on the voltage noise fields $\{\delta V_{g,k}(t);k=1,...,N\}$ (see Appendix~\ref{App_A}) and satisfy the independent random Gaussian processes, i.e., $\Braket{\eta_{k}(t)}_{s}=0$ and the correlation function
\begin{equation}
K_{k,k'}(t,t')=\Braket{\eta_{k}(t)\eta_{k'}(t')}_{s}.
\end{equation}
Here $\Braket{\cdots}_{s}$ denotes averaging over stochastic realizations. The characteristic frequency $\Gamma$ measures the noise intensity at $\pi f_{0}$. In this work, we choose $\Gamma\ll\epsilon$, i.e., the weak quantum-system-environment interaction. We also set $\epsilon/\Gamma=10$, $2f_{0}/\Gamma=10^{3}$ and $2\pi f_{0}>\lambda,\epsilon$. The power spectral density (PSD) of the noise is calculated by
\begin{equation}
{\cal S}_{\eta}(2\pi f)=\int^{\infty}_{-\infty}K_{k,k}(t,t+\tau)e^{-i2\pi f\tau}d\tau.
\end{equation}
For the white noise, $K_{k,k'}(t,t')=\delta_{k,k'}\delta(t-t')$ and ${\cal S}_{\eta}(2\pi f)$ is independent of the noise frequency $f$, i.e., ${\cal S}_{\eta}(2\pi f)\propto f^{0}$.

The generation of colored noise fields may be implemented by digitally filtering a stochastic white field via four steps~\cite{RJAV:Zhivomirov2018}: (i) A continuous white-field signal is discretized into a sequence; (ii) This finite discrete sequence is then converted into a same-length complex-valued sequence via the discrete Fourier transform; (iii) Each Fourier component in the sequence is multiplied by a frequency-dependent factor; and (iv) Finally, the colored noise sequence in the time domain is given by the inverse discrete Fourier transform. More detailed algorithm can be found in Appendix~\ref{App_B}.

Figure~\ref{Fig1}b depicts the dependence of $K_{k,k'}(t,t')$ on the temporal difference $(t-t')$ for pink- and blue-colored noise fields, whose PSDs are ${\cal S}_{\eta}(2\pi f)\propto f^{-1}$ and $\propto f^{1}$ (see Fig.~\ref{Fig5}c). Here different colored noise fields have the same PSD value at the Nyquist frequency $\pi f_{0}$. The result $K_{k,k'\neq k}(t,t')=0$ denotes that two noise processes are independent (uncorrelated). The pink-colored $K_{k,k}(t,t')$ is positive at any time, indicating a predictable relationship in the co-occurrence of two events in a process. In contrast, the blue-colored $K_{k,k}(t,t')$ is negative at most of the time, meaning two events are unlikely to occur together.

In the absence of noise fields $\eta_{k}(t)$, the relaxation ($T_{1}$) and dephasing ($T_{2}$) times of individual qubit are limited down to $\sim10$ ns~\cite{Science:Devoret2013} because of the inevitable coupling to the local electromagnetic (EM) fluctuation~\cite{PRL:Astafiev2004}. The maximum timescale of interest in this work $t_{\textrm{max}}$ is set to be much shorter than $T_{1,2}$, i.e., $t_{\textrm{max}}=10\Gamma^{-1}\ll T_{1,2}$. In addition, $T_{1,2}$ may be extended by reducing the sensitivity of the qubit to the local EM reservoir~\cite{PRA:Koch2007} or feedback-controlling the qubit dynamics~\cite{NJP:Yu2018,PRA:Yu2018}. The qubit-local-reservoir interaction thereby is neglected.

\begin{figure}
\includegraphics[width=5.5cm]{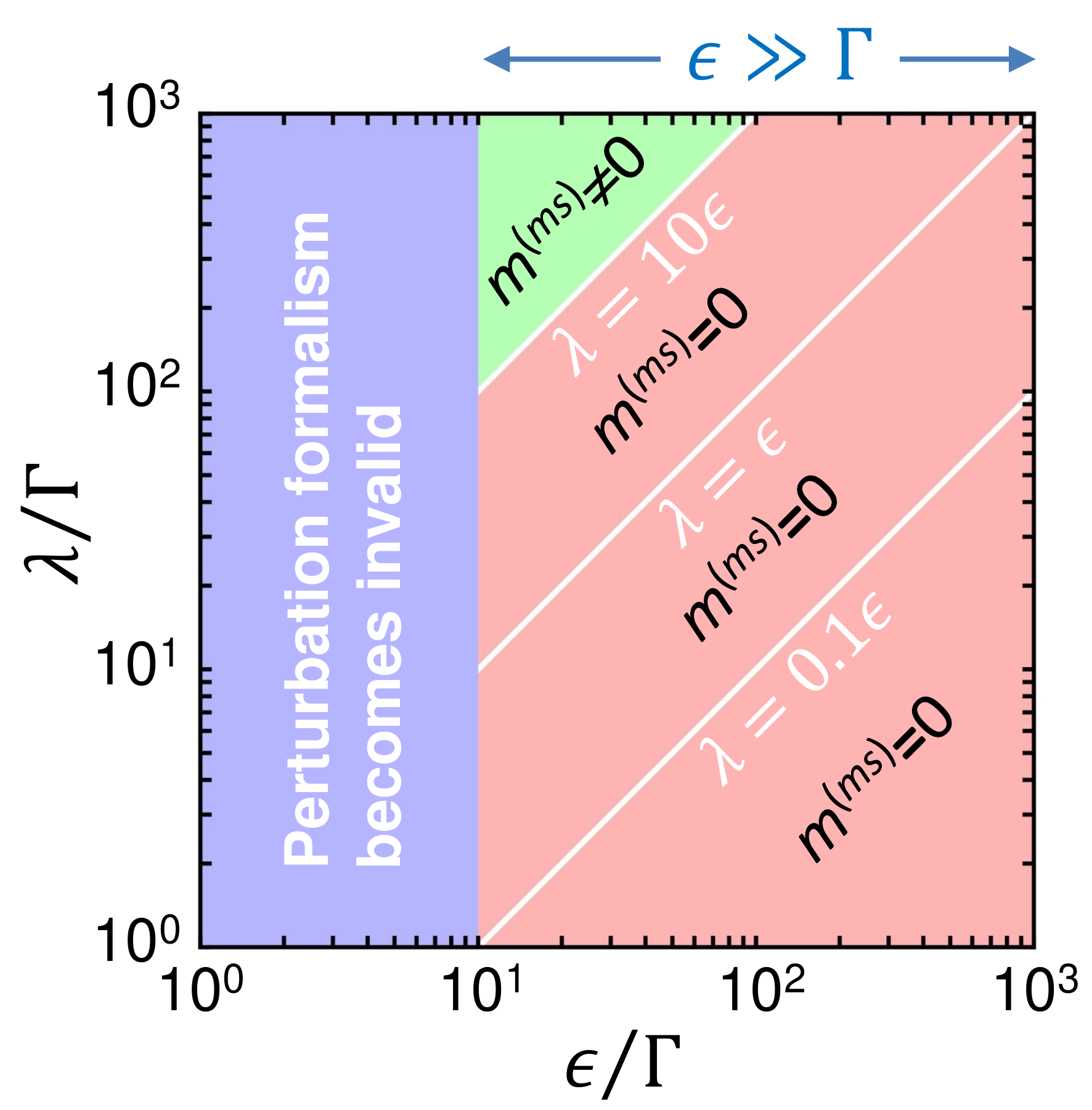}\\
\caption{(Color online) Parameters $\epsilon$ and $\lambda$ of Ising model. Non-Markovian master equation~(\ref{MasterEq1}) is only valid in the regime of $\epsilon\gg\Gamma$, where the noise fields $\eta_{k}(t)$ with the intensity $\Gamma$ at the Nyquist frequency $\pi f_{0}$ perturb the spin states. This valid $\epsilon-\lambda$ area may be further divided into two subareas, respectively represented by green and red color, according to the ratio $\lambda/\epsilon$. Generally, in the strong-interspin-interaction regime, i.e., $\lambda/\epsilon\gg1$, the metastable value of magnetization $m^{(ms)}$ is nonzero while $m^{(ms)}=0$ in the rest of valid regime. In this work, the characteristic frequency $\Gamma$ is much larger than the common relaxation $T^{-1}_{1}$ and dephasing $T^{-1}_{2}$ rates of SC qubit and the maximum time scale of interest $t_{\textrm{max}}$ is much shorter than $T_{1,2}$. The Nyquist frequency $\pi f_{0}$ is set to be larger than both $\epsilon$ and $\lambda$.}\label{Fig2}
\end{figure}

According to the perturbation formalism in~\cite{PRL:Chenu2017}, the dissipative process of the system is described by the non-Markovian master equation for the density matrix $\rho(t)$
\begin{eqnarray}\label{MasterEq1}
\nonumber\frac{d}{dt}\rho(t)&=&-\frac{i}{\hbar}[H_{s},\rho(t)]-\sum_{k,k'}\frac{\Gamma}{2}\int^{t}_{0}K_{k,k'}(t,t')\\
&&\times[\sigma^{x}_{k},[U_{s}(t,t')\sigma^{x}_{k'}U^{\dag}_{s}(t,t'),\rho(t)]]dt',
\end{eqnarray}
with the time-evolution operator
\begin{equation}
U_{s}(t,t')={\cal{T}}\exp[-\frac{i}{\hbar}\int^{t}_{t'}H_{s}(\tau)d\tau],
\end{equation}
and the time-ordering operator ${\cal{T}}$. In deriving Eq.~(\ref{MasterEq1}), we have applied the assumption $\epsilon\gg\Gamma$ (see Fig.~\ref{Fig2}), i.e., the intensity $\Gamma$ of noise fields $\eta_{k}(t)$ is weak enough that the noise fields only perturb the spin states. When the noise fields become strong, the second term on the right side of Hamiltonian~(\ref{Hamiltonian1}) is comparable to $H_{s}$ in energy and Eq.~(\ref{MasterEq1}) is no longer valid. The non-Markovian master equation~(\ref{MasterEq1}) may be written in the local-in-time form (see Sec.~\ref{Introduction}), though a time convolution integral complicates the calculation.

Solving Eq.~(\ref{MasterEq1}) relies on the specific noise spectrum. As we will see below, for the memoryless white noise, the conventional approaches, such as direct solving the matrix differential equation and the Monte Carlo wave-function (MCWF) method~\cite{JOSAB:Molmer1993}, are applicable. In contrast, for the colored noise fields, direct solving the matrix differential equation is more favorable. In the following, we focus on the quantum ensemble average of the single-spin magnetization observable
\begin{equation}
m(t)=N^{-1}\sum_{k}\Braket{\sigma_{k}^{z}(t)}_{e},
\end{equation}
with the definition $\Braket{...}_{e}=\Tr[\rho(t)...]$. An arbitrary noise source can be decomposed into a series of independent colored terms. Our aim is to explore the relaxation evolution of $m(t)$ responding to different colored components. We will focus on three typical noise fields: white, red- and blue-colored noise fields.

\begin{figure*}
\includegraphics[width=18cm]{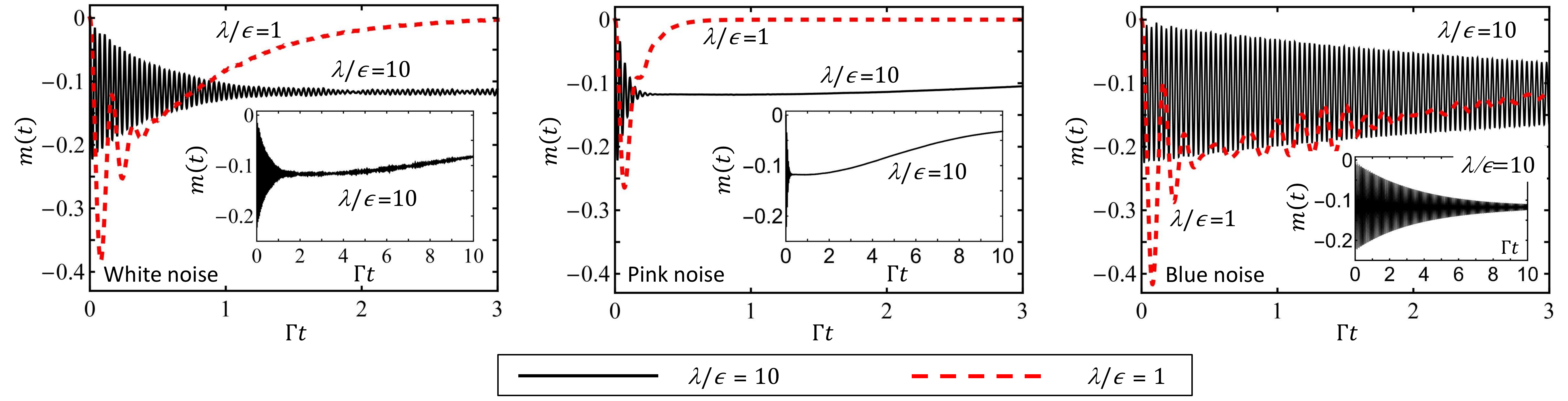}\\
\caption{(Color online) Quantum Ising model perturbed by white, red- and blue-colored noise fields. Relaxation dynamics of magnetization $m(t)$ with $\lambda/\epsilon=1$ and 10 for white (left), red-colored (middle) and blue-colored (right) noise fields. For all curves, the system is initialized in the unpolarized state $\prod_{k}\otimes(\Ket{\uparrow}-\Ket{\downarrow})_{k}=\prod_{k}\otimes\Ket{0}_{k}$. The insets show $m(t)$ within a long-time duration for $\lambda/\epsilon=10$.}\label{Fig3}
\end{figure*}

\section{White Noise}

We start with the quantum many-body system coupled to the white reservoir, where the noise fields $\eta_{k}(t)$ follow the memoryless processes with $K_{k,k'}(t,t')=\delta_{k,k'}\delta(t-t')$. In this limit, the master equation~(\ref{MasterEq1}) returns to the Markovian form
\begin{equation}\label{MarkovianMasterEq}
\frac{d}{dt}\rho(t)={\cal{L}}(\rho),
\end{equation}
with the dynamical generator
\begin{equation}
\nonumber{\cal{L}}(\rho)=-\frac{i}{\hbar}[H_{s},\rho]-\frac{\Gamma}{2}\sum_{k}(\sigma^{x}_{k}\sigma^{x}_{k}\rho+\rho\sigma^{x}_{k}\sigma^{x}_{k}-2\sigma^{x}_{k}\rho\sigma^{x}_{k}).
\end{equation}
The quantum jump (also called Lindblad) operators~\cite{JOSAB:Molmer1993}
\begin{equation}\label{QuantumJumpOperator}
\sqrt{\Gamma}\sigma^{x}_{k}=\sqrt{\Gamma}\sigma^{-}_{k}+\sqrt{\Gamma}(\sigma^{-}_{k})^{\dag},
\end{equation}
denote that the incoherent lowering $\sigma^{-}_{k}=(\Ket{\downarrow}\Bra{\uparrow})_{k}$ and raising $(\sigma^{-}_{k})^{\dag}$ of the $\Ket{\uparrow}$-state population occur at the same rate $\Gamma$.

In general, the mean-field analysis, i.e., neglecting the interspin correlation $\Braket{\sigma^{x}_{k}\sigma^{x}_{k'}}_{e}\simeq\Braket{\sigma^{x}_{k}}_{e}\Braket{\sigma^{x}_{k'}}_{e}$ for $k\neq k'$, may provide an instructive insight in the relevant physical mechanisms. Defining $x=\Braket{\sigma^{x}_{k}}_{e}$, $y=\Braket{\sigma^{y}_{k}}_{e}$ and $m=\Braket{\sigma^{z}_{k}}_{e}$, we obtain
\begin{eqnarray}
\frac{d}{dt}x&=&2\epsilon y,\\
\frac{d}{dt}y&=&-2\Gamma y-2\epsilon x-2\lambda xm,\\
\frac{d}{dt}m&=&-2\Gamma m+2\lambda xy.
\end{eqnarray}
Setting $\frac{d}{dt}x=\frac{d}{dt}y=\frac{d}{dt}m=0$ leads to the trivial steady-state ($ss$) solutions when $t\rightarrow\infty$ (a time scale much longer than $\Gamma^{-1}$),
\begin{equation}
x^{(ss)}=y^{(ss)}=m^{(ss)}=0.
\end{equation}
This is unlike the typical driven-dissipative many-body systems composed of interacting atoms~\cite{PRL:Olmos2013} or photons~\cite{PRL:Foss-Feig2017}, where the mean-field treatment predicts a dynamical first-order phase transition. Thus, studying this open quantum system relies on numerically solving the master equation~(\ref{MarkovianMasterEq}).

\begin{figure*}
\includegraphics[width=18cm]{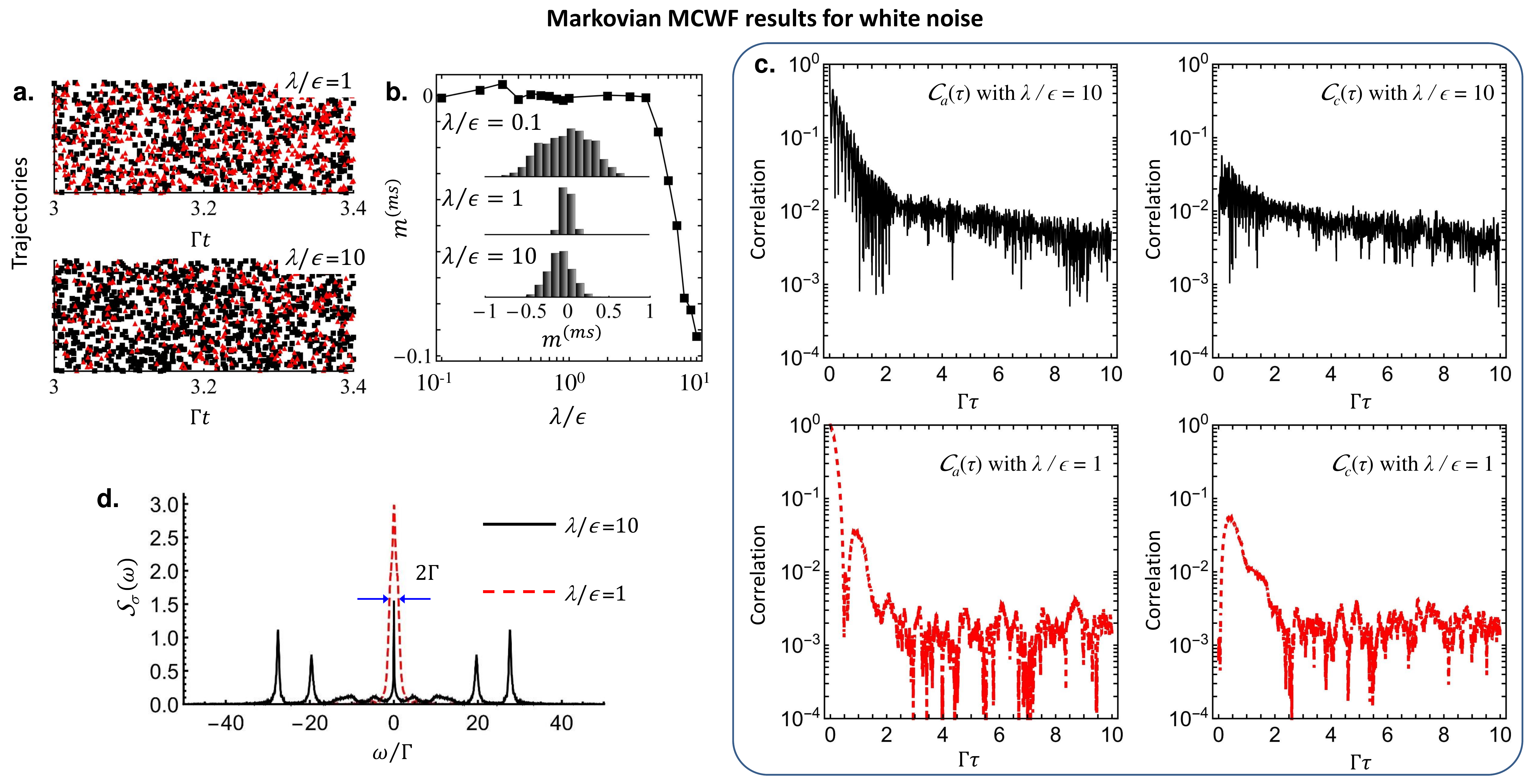}\\
\caption{(Color online) Results derived from MCWF method for white-noise-perturbed system. (a) Quantum jumps recorded in an ensemble of trajectories. Filled-square markers correspond to the incoherent raising events ($\Ket{\downarrow}\rightarrow\Ket{\uparrow}$), i.e., the jumps increase $m(t)$, while filled-triangle markers denote the incoherent lowering events ($\Ket{\uparrow}\rightarrow\Ket{\downarrow}$), i.e., $m(t)$ is reduced after the jumps. (b) Metastable value $m^{ms}=m(t_{\textrm{max}})$ as a function of $\lambda/\epsilon$. As the system enters the strong interaction regime $\lambda/\epsilon\gg1$, $m^{ms}$ becomes negative and $|m^{ms}|$ is increased. The inset lists statistical distribution of trajectories of $m^{ms}$ for several different $\lambda/\epsilon$. (c) Auto- and cross-correlation functions ${\cal C}_{a,c}(\tau)$ with $\lambda/\epsilon=1$ and 10. The corresponding power spectral densities ${\cal S}_{\sigma}(\omega)$ are shown in (d).}\label{Fig4}
\end{figure*}

Exploring many-body dissipative system in the thermodynamic limit $N\rightarrow\infty$ is mathematically impractical. The finite amount of computer memory and computational time limit the solvable system size. Here we consider the system with $N=8$ to capture the general features of larger systems. In general, the master equation~(\ref{MarkovianMasterEq}) for a small system may be solved numerically via the exact diagonalization method~\cite{JModOpt:Yu2016}, the approach of direct solving the matrix differential equation and the Monte Carlo wave-function (MCWF) method~\cite{JOSAB:Molmer1993}. In comparison, the MCWF approach, where the density matrix $\rho$ is treated as an ensemble of state vectors suffered irreversible quantum jumps, can also be applied to compute two-time correlation functions. Here we employ the MCWF method to simulate the time evolution of $m(t)$ in the Markovian limit. The main results are summarized in Fig.~\ref{Fig3} (left) and~\ref{Fig4}.

As shown in Fig.~\ref{Fig3} (left), in the medium $\lambda/\epsilon\sim1$ (also, weak $\lambda/\epsilon\sim0$) interaction regime, $m(t)$ approaches the stationary state $m^{ss}=0$ after a timescale of $\Gamma^{-1}$. In the steady state, the number of quantum jumps reducing the $\Ket{\uparrow}$-population matches that of the jumps increasing the $\Ket{\uparrow}$-population (see Fig.~\ref{Fig4}a), i.e., reaching a balance between incoherent lowering and raising events. Indeed, the number of incoherent lowering (raising) events is given by the $\Ket{\uparrow}$-population ($\Ket{\downarrow}$-population) times the corresponding quantum-jump rate. Since the incoherent lowering and raising quantum-jump rates are both equal to $\Gamma$ [see Eq.~(\ref{QuantumJumpOperator})], the populations in $\Ket{\downarrow}$ and $\Ket{\uparrow}$ states are same, leading to $m^{ss}=0$. In contrast, for the system in the strong coupling regime ($\lambda/\epsilon\gg1$), $m(t)$ relaxes rapidly to states with metastable ($ms$) characteristics where $m^{ms}\neq0$ and then decays slowly to the true stationary state $m^{ss}$. Interestingly, during the metastable period the number of incoherent-raising events is apparently larger than that of incoherent-lowering events (see Fig.~\ref{Fig4}a), which indicates that the $\Ket{\downarrow}$-population must be higher than the $\Ket{\uparrow}$-population, resulting in $m^{ms}<0$ (see Fig.~\ref{Fig4}b). This metastability may be interpreted from the spectral structure of the generator ${\cal{L}}$, where a large gap separates a few eigenvalues with the real parts equal or close to zero from the others~\cite{PRL:Macieszczak2016} (see Appendix~\ref{App_B}).

To quantitatively evaluate the metastability of the dissipative system, we define the single-spin magnetization at the maximum timescale of interest in this work as the metastable value $m^{ms}$,
\begin{equation}
m^{ms}=m(t_{\textrm{max}}).
\end{equation}
The dependence of $m^{ms}$ on the spin-spin interaction $\lambda$ is plotted in Fig.~\ref{Fig4}b. As $\lambda/\epsilon$ is raised from zero, the mean value of the trajectory ensemble of $m^{ms}$ stays at zero, i.e., non-metastability, while the spread of the statistical distribution of $m^{ms}$ is strongly narrowed, indicating the build up of interspin correlations. When $\lambda/\epsilon$ is further increased, the ensemble average of $m^{ms}$ starts to be negative and $|m^{ms}|$ grows fast. However, the trajectory-distribution width of $m^{ms}$ becomes larger because the quantum jumps cause the strong fluctuation in the enhanced spin-spin interaction term in $H_{s}$. Generally, the metastable behavior $m^{ms}\neq0$ for the open quantum system in the Markovian limit occurs in the strong interaction regime (see Fig.~\ref{Fig2}).

By means of the MCWF method, we further consider auto- and cross-correlation functions
\begin{eqnarray}
{\cal C}_{a}(\tau)&=&N^{-1}\sum_{k}\Braket{\sigma^{z}_{k}(t+\tau)\sigma^{z}_{k}(t)}_{e},\\
{\cal C}_{c}(\tau)&=&[N(N-1)]^{-1}\sum_{k\neq k'}\Braket{\sigma^{z}_{k}(t+\tau)\sigma^{z}_{k'}(t)}_{e}.
\end{eqnarray}
The term $\sigma^{z}_{k}(t)$ denotes the operator $\sigma^{z}_{k}$ in the Heisenberg picture. The PSD of single spin's relaxation dynamics is given by the Fourier transform of ${\cal C}_{a}(\tau)$,
\begin{equation}
{\cal S}_{\sigma}(\omega)=\int^{\infty}_{-\infty}{\cal C}_{a}(\tau)e^{-i\omega t}dt,
\end{equation}
while ${\cal C}_{c}(\tau)$ measures the similarity between dynamics of two spins. As illustrated in Fig.~\ref{Fig4}c, the spin-spin interaction $\lambda$ strongly controls the behaviors of ${\cal C}_{a,c}(\tau)$. For a small $\lambda$, e.g., $\lambda/\epsilon=1$, ${\cal C}_{a,c}(\tau)$ are mostly overlapped with each other and decay to zero at a rate of $\Gamma$. As $\lambda$ is increased, ${\cal C}_{a,c}(\tau)$ are much enhanced over the time duration of interest. For $\lambda/\epsilon\gg1$, ${\cal C}_{a}(\tau)$ displays two distinct temporal regimes. When $\tau<(\ln10)/\Gamma$, ${\cal C}_{a}(\tau)$ is higher than ${\cal C}_{c}(\tau)$ and decays at a rate of $\Gamma/(\ln10)$. In contrast, after $(\ln10)/\Gamma$ the decay of ${\cal C}_{a}(\tau)$ is much slowed down and overlapped with ${\cal C}_{c}(\tau)$.

The inset of Fig.~\ref{Fig4}d display two examples of the power spectral density ${\cal S}_{\sigma}(\omega)$ of the quantum system coupling to the white environment. In the weak $\lambda/\epsilon\sim0$ and medium $\lambda/\epsilon\sim1$ interaction regimes, a single peak is located at the central position with the full width at half maximum (FWHM) of $\sim\Gamma$, which denotes that the system relaxation timescale approximates $\Gamma^{-1}$. In contrast, ${\cal S}_{\sigma}(\omega)$ of the strong coupling system ($\lambda/\epsilon\gg1$) exhibits two features: ($i$) The FWHM of the central peak is much smaller than $\Gamma$, indicating the existence of metastable state; and ($ii$) Multiple side peaks split off the central peak, which is similar to the Mollow triplet spectrum in quantum optics~\cite{PR:Mollow1969}. Indeed, this multiplet lineshape arises from the spin-spin interaction term in $H_{s}$ modulating the system's relaxation dynamics. Since the interspin coupling can only induce the transition between two states with the $\ket{\uparrow}$-number difference of $0$ or $2n$ with $n\in\mathbb{N}$, the positions of side peaks are estimated to be $\pm(\frac{\lambda}{4N}-2\epsilon)$ and $\pm 2n(\frac{\lambda}{4N}-2\epsilon)$.

\section{Colored Noises}

So far we have only considered the quantum system perturbed by the memoryless white noise field. For the colored perturbations, the effect of memorizing historical events may lead to a different time evolution of the dissipative system for a given initial state. The Markovian master equation~(\ref{MarkovianMasterEq}) is no longer valid and one has to solve the non-Markovian master equation~(\ref{MasterEq1}). We choose the eigenstates $\{\Ket{\alpha};\alpha=0,1,...,2^{N}\}$ of $H_{s}$ to span the Hilbert space, i.e., $H_{s}\Ket{\alpha}=\hbar\omega_{\alpha}\Ket{\alpha}$ with the corresponding eigenvalue $\omega_{\alpha}$ and $\omega_{\alpha}\leq\omega_{\alpha+1}$. Equation~(\ref{MasterEq1}) is then rewritten as
\begin{eqnarray}\label{MasterEq2}
\nonumber\frac{d}{dt}\rho(t)&=&-\frac{i}{\hbar}[H_{s},\rho(t)]-\sum_{k}\sum_{\alpha,\alpha'}\frac{\Gamma}{2}\\
&&\times[\sigma^{x}_{k},[\Bra{\alpha}\sigma^{x}_{k}\Ket{\alpha'}{\cal{K}}^{\alpha\alpha'}_{k}(t)\Ket{\alpha}\Bra{\alpha'},\rho(t)]],~~~~~
\end{eqnarray}
where ${\cal{K}}^{\alpha\alpha'}_{k}(t)=\int^{t}_{0}K_{k,k}(t,t')e^{-i\Delta\omega_{\alpha\alpha'}t'}dt'$ with $\Delta\omega_{\alpha\alpha'}=\omega_{\alpha}-\omega_{\alpha'}$, under the new basis. Here we have also used the fact that $K_{k,k'}(t,t')=0$ for $k\neq k'$.

Equation~(\ref{MasterEq2}) can be expressed in the local-in-time form (see Sec.~\ref{Introduction}). The earlier numerical treatment on the local-in-time master equation relies on extending the Hilbert space of the quantum system~\cite{PRA:Breuer1999} or the stochastic system state evolution conditioned on the environment hidden variable~\cite{PRA:Gambetta2003}. However, the needed large computer memory restricts the system size that can be studied. In addition, the time convolution in Eq.~(\ref{MasterEq2}) makes it difficult to utilize the improved MCWF method developed in~\cite{PRL:Piilo2008}. Thus, we directly compute the non-Markovian Eq.~(\ref{MasterEq2}) via treating it as a differential matrix equation. This approach is not applicable to derive two-time correlation functions. Here we only consider the single-spin magnetization $m(t)$.

Figure~\ref{Fig3} (middle) depicts the time-evolution examples of the pink-colored-noise-perturbed quantum system. The system relaxation is apparently accelerated in comparison with the Markovian results shown in Fig.~\ref{Fig3} (left). Even in the strong coupling regime $\lambda/\epsilon\gg1$, the metastable period is shortened and the dissipative system nearly arrives at the true stationary state $m^{(ss)}=0$ within $t_{\textrm{max}}$. The reason lies in the fact that $K_{k,k}(t,t')>0$ for the pink-colored noise field (see Fig.~\ref{Fig1}b). To give an intuitive explanation, we make a rough approximation by neglecting the difference among the sampling components in ${\cal{K}}^{\alpha\alpha'}_{k}(t)$, i.e., replacing $e^{-i\Delta\omega_{\alpha\alpha'}t}$ by 1. Then, the master equation is further simplified as
\begin{equation}
\frac{d}{dt}\rho(t)=-\frac{i}{\hbar}[H_{s},\rho(t)]-\sum_{k}\frac{\tilde{\Gamma}_{k}(t)}{2}[\sigma^{x}_{k},[\sigma^{x}_{k},\rho(t)]],
\end{equation}
where the memory of the colored noise history is involved in the effective characteristic decay rate $\tilde{\Gamma}_{k}(t)={\cal{K}}_{k}(t)\Gamma$ with the time-dependent enhancement factor ${\cal{K}}_{k}(t)=\int^{t}_{0}K_{k,k}(t,t')dt'$. The positive memory of the pink-colored noise leads to ${\cal{K}}_{k}(t)>1$, thereby accelerating the system relaxation. In contrast, the relaxation of the quantum system interacting with the blue-colored reservoir is slowed down compared to the Markovian results [see Fig.~\ref{Fig3} (right)]. Especially in the strong coupling limit ($\lambda/\epsilon\gg1$), the system does not even reach the metastable state within $t_{\textrm{max}}$. Again, following the rough approximation, the negative memory of the blue noise leads to $0<{\cal{K}}_{k}(t)<1$ (see Fig.~\ref{Fig1}b) and $\tilde{\Gamma}_{k}(t)<\Gamma$.

Such an acceleration or slowing down of the system relaxation may also be interpreted from the noise PSDs (see Fig.~\ref{Fig1}c). All transition frequencies $|\Delta\omega_{\alpha\alpha'}|$ are smaller than the Nyquist frequency $\pi f_{0}$ due to $\pi f_{0}>\epsilon$ and $\lambda$. Since the PSD of pink-colored (blue-colored) noise is higher (lower) than that of white noise for $f<f_{0}$, the colored-noise perturbation on the quantum system is enhanced (weakened).

\begin{figure}
\includegraphics[width=7.0cm]{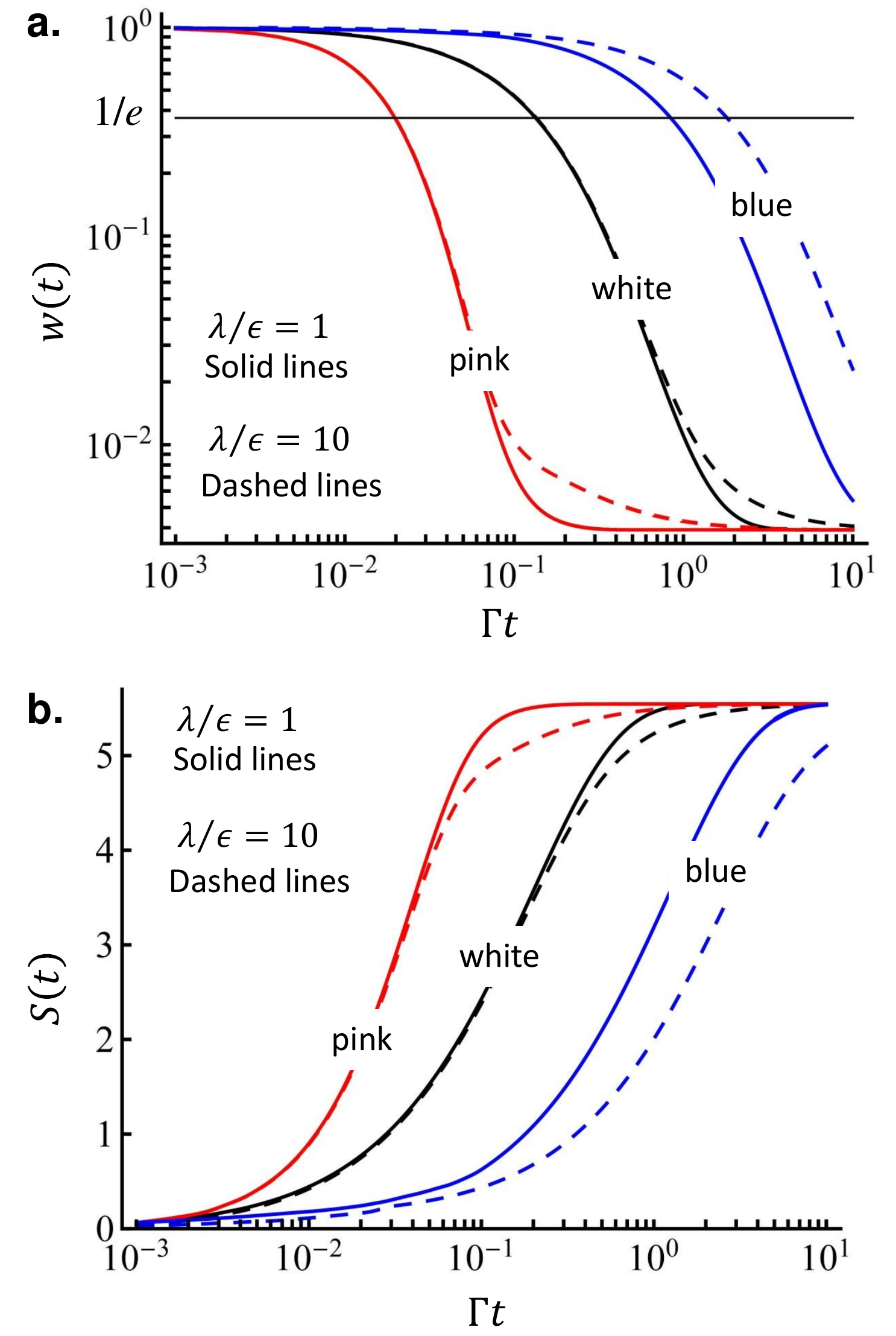}\
\caption{(Color online) Ground state of many-body quantum system. Ground-state weight $w(t)$ (a) and entropy $S(t)$ (b) for the system perturbed by the white, red-colored and blue-colored noise fields. The solid and dashed lines correspond to $\lambda/\epsilon=1$ and $10$, respectively. For all curves, the system is initially prepared in the ground state $\Ket{\alpha=0}$.}\label{Fig5}
\end{figure}

We further consider the lifetime of the many-body system's ground state $\Ket{\alpha=0}$. In the limit of $\lambda/\epsilon\ll1$, all qubits are close to independent and $\Ket{\alpha=0}$ approximates the fully-polarized state $\Pi_{k}\otimes\Ket{\uparrow}_{k}$. For $\lambda/\epsilon\gg1$, one has $\bra{\alpha=0}(\sum_{k}\sigma^{x}_{k})^{2}\Ket{\alpha=0}\simeq0$. We assume that the system is initialized at $\Ket{\alpha=0}$. The dissipation operators $\{\sigma^{x}_{k};k=1,...,N\}$ mix $\Ket{\alpha=0}$ with $\Ket{\alpha>0}$, reducing its weight in the wavefunction
\begin{equation}
w(t)=\Bra{\alpha=0}\rho(t)\Ket{\alpha=0}.
\end{equation}
The ground-state lifetime $\tau_{G}$ is defined as the timescale at which $w(t)$ decreases to $1/e$. As displayed in Fig.~\ref{Fig5}a, $\tau_{G}$ is shortened (extended) for the quantum system interacting with the pink-colored (blue-colored) reservoir in comparison with that of the white-noise-perturbed quantum system. Interestingly, Fig.~\ref{Fig5}a illustrates that $\tau_{G}$ rarely depends on the spin-spin interaction $\lambda$, especially for the white and pink-colored noise fields. This is because the metastable behavior of the system also relies on the initial state of the system (see Appendix~\ref{App_B}). It is unfair to compare the lifetimes of two different (initial) ground states. The von Neumann entropy
\begin{equation}
S=-\Tr(\rho\ln\rho),
\end{equation}
shown in Fig.~\ref{Fig5}b is commonly employed to quantify the departure of the system from a pure state. $S$ starts from zero because $\Ket{\alpha=0}$ is also an eigenstate of $\rho$ at $t=0$. Then, $S$ rises sharply, indicating that the system rapidly evolves to a mixed state. Finally, $S$ asymptotically approaches the maximum value $\ln2^{N}$. In comparison with the white noise, the slope of $S$ vs. $t$ raises (degrades) for the pink-colored (blue-colored) noise.

\section{Summary}

In conclusion, we have studied the dissipative Ising model based on a SC-circuit platform. The exceptional flexibility of quantum circuits enables designing of quantum jump operators and arbitrary tailoring of the environmental noise, whose implementations are not straightforward in quantum gaseous or photonic platforms. Unlike the common Ising system with short-range (e.g., nearest-neighbor and next-nearest-neighbor) interparticle interactions, each qubit is equally coupled to others in the many-body system. We focus on the relaxation of magnetization observable $m(t)$ which is numerically simulated based on the non-Markovian master equation in the perturbation approximation~\cite{PRL:Chenu2017}.

For the quantum system coupled to the memoryless white reservoir, the metastable behavior ($m^{(ms)}\neq0$) unaccompanied by the first-order phase transition is predicted in the strong spin--spin coupling regime ($\lambda/\epsilon\gg1$). The corresponding PSD of the system's dissipative dynamics exhibits multiple peaks, among which the central one possesses a narrow linewidth ($\ll\Gamma$). This metastability arises entirely from the strong interspin interaction. The system relaxation in two specific colored, pink and blue, noise fields are compared with the white-noise case. It is found that the pink-colored perturbation accelerates the system approaching the true stationary state while the relaxation process of the system becomes slow in the blue-colored environment. This can be understood from the correlation functions $K_{k,k}(t,t')$ of different colored noise fields. For the pink-colored noise field, $K_{k,k}(t,t')$ is positive at any time difference $(t-t')$, enhancing the effective characteristic decay rate. By contrast, $K_{k,k}(t,t')$ of the blue-colored noise stays at a negative value except when $(t-t')=0$, reducing the system's effective decay rate. Following the similar way, one may study open quantum systems perturbed by other colored noise fields.


\begin{acknowledgments}
This research has been supported by the National Research Foundation Singapore \& by the Ministry of Education Singapore Academic Research Fund Tier 1 (Grant No. 2016-T1-002-032-01).
\end{acknowledgments}

\appendix 

\section{Hamiltonian of many-body system $H$}\label{App_A}

The amount of charge $Q_{i,k}$ in the $k$-th Cooper-pair box is given by
\begin{equation}
Q_{i,k}=Q_{g,k}+Q_{j,k}+Q_{c,k},
\end{equation}
where $Q_{g,k}$, $Q_{j,k}$ and $Q_{c,k}$ represent respectively the charges on the plates of gate capacitor $C_{g}$, self-capacitor $C_{j}$ and coupling capacitor $C_{c}$ that are linked to the box. The gate voltage is then equal to
\begin{equation}
V_{g,k}=-\frac{Q_{g,k}}{C_{g}}+\frac{Q_{j,k}}{C_{j}}.
\end{equation}
In addition, the electric charge conversion leads to
\begin{equation}
\sum_{k}Q_{c,k}=0,
\end{equation}
and we further have
\begin{equation}
-\frac{Q_{c,1}}{C_{c}}+\frac{Q_{j,1}}{C_{j}}=-\frac{Q_{c,2}}{C_{c}}+\frac{Q_{j,2}}{C_{j}}=\cdots.
\end{equation}
The charges $Q_{g,k}$, $Q_{j,k}$ and $Q_{c,k}$ may be expressed in terms of $Q_{i,k}$ and $V_{g,k}$,
\begin{eqnarray}
Q_{g,k}&=&\frac{C_{g}}{C_{\Sigma}}(C_{g}V_{g,k}+Q_{i,k}+C_{c}V_{c})-C_{g}V_{g,k},\\
Q_{j,k}&=&\frac{C_{j}}{C_{\Sigma}}(C_{g}V_{g,k}+Q_{i,k})+\frac{C_{j}}{C_{\Sigma}}C_{c}V_{c},\\
Q_{c,k}&=&\frac{C_{c}}{C_{\Sigma}}(C_{g}V_{g,k}+Q_{i,k})-\frac{C_{g}+C_{j}}{C_{\Sigma}}C_{c}V_{c},
\end{eqnarray}
where $V_{c}$ is defined as $V_{c}=\frac{\sum_{k}(C_{g}V_{g,k}+Q_{i,k})}{N(C_{g}+C_{j})}$. Combining the total electrostatic and tunneling energies of Cooper pairs in the boxes, one obtains the Hamiltonian of multi-charge-qubit system
\begin{eqnarray}
\nonumber H&=&\sum_{k}[E_{C}(N_{g,k}-N_{k})^{2}-E_{J}\cos\phi_{k}]\\
&&+\left[\sum_{k}\frac{E_{C}}{\sqrt{NV}}(N_{g,k}-N_{k})\right]^{2}.
\end{eqnarray}
The operator $\phi_{k}$ corresponds to the phase difference of Cooper pairs across the $k$-th junction. The operator $N_{k}=-\frac{Q_{i,k}}{2e}$ counts the number of Cooper pairs in the $k$-th box. The operator $N_{g,k}=\frac{C_{g}V_{g,k}}{2e}$ is the gate-charge bias. The ratio $\frac{E^{2}_{C}}{V}$ with $V=\frac{(2e)^{2}}{2C_{c}}(1-\frac{C_{c}}{C_{\Sigma}})$ measures the interqubit coupling. In the charge-number representation, the operators $N_{k}$ and $\cos\phi_{k}$ are written as
\begin{eqnarray}
N_{k}&=&\sum_{n_{k}}n_{k}\Ket{n_{k}}\Bra{n_{k}},\\
\cos\phi_{k}&=&\frac{1}{2}\sum_{n_{k}}(\Ket{n_{k}}\Bra{n_{k}+1}+\Ket{n_{k}+1}\Bra{n_{k}}).
\end{eqnarray}
where $n_{k}\in\mathbb{Z}$ denotes the number of excess Cooper pairs in the box. We divide the gate voltage $V_{g,k}$ into two parts, $V_{g,k}=V_{0}+\delta V_{g,k}(t)$, for which the gate charge bias is re-expressed as $N_{g,k}=(N_{0}+\frac{1}{2})-\frac{\delta N_{g,k}(t)}{2}=\frac{C_{g}V_{0}}{2e}+\frac{C_{g}\delta V_{g,k}(t)}{2e}$ with $N_{0}\in\mathbb{Z}$. $\delta V_{g,k}(t)$ is the external voltage noise and $\delta N_{g,k}(t)$ is the corresponding gate-charge-bias fluctuation.

In the two-state approximation, the Hamiltonian $H$ is simplified as
\begin{eqnarray}
\nonumber H&=&\sum_{k}\left[\delta N_{g,k}(t)+\frac{E_{C}}{NV}\sum_{k'}\delta N_{g,k'}(t)\right]\frac{E_{C}}{2}s^{z}_{k}\\
&&-\sum_{k}\frac{E_{J}}{2}s^{x}_{k}+\frac{E^{2}_{C}}{2NV}\sum_{k<k'}s^{z}_{k}s^{z}_{k'},
\end{eqnarray}
with the operators $s^{x}_{k}=(\Ket{1}\Bra{0})_{k}+(\Ket{0}\Bra{1})_{k}$ and $s^{z}_{k}=(\Ket{1}\Bra{1})_{k}-(\Ket{0}\Bra{0})_{k}$. Defining two spin states $\Ket{\uparrow}_{k}=\frac{1}{\sqrt{2}}(\Ket{1}_{k}+\Ket{0}_{k})$ and $\Ket{\downarrow}_{k}=\frac{1}{\sqrt{2}}(\Ket{1}_{k}-\Ket{0}_{k})$, $s^{x,z}_{k}$ should be replaced by $\sigma^{z,x}_{k}$, and we arrive at
\begin{eqnarray}
\nonumber H&=&-\frac{E_{J}}{2}\sum_{k}\sigma^{z}_{k}+\frac{E^{2}_{C}}{2NV}\sum_{k<k'}\sigma^{x}_{k}\sigma^{x}_{k'}\\
&&+\frac{E_{C}}{2}\sum_{k}\left[\delta N_{g,k}(t)+\frac{E_{C}}{NV}\sum_{k'}\delta N_{g,k'}(t)\right]\sigma^{x}_{k}.~~~~~~~
\end{eqnarray}
Defining $\hbar\epsilon=\frac{E_{J}}{2}$, $\hbar\lambda=\frac{E^{2}_{C}}{2V}$ and $\hbar\sqrt{\frac{\Gamma}{2}}\eta_{k}(t)=\frac{E_{C}}{2}[\delta N_{g,k}(t)+\frac{E_{C}}{NV}\sum_{k'}\delta N_{g,k'}(t)]$, the Hamiltonian $H$ can be re-written as the form shown in the main text. For a large $N$, one has $\sum_{k'}\delta N_{g,k'}(t)\simeq0$ and $\hbar\sqrt{\frac{\Gamma}{2}}\eta_{k}(t)\simeq\frac{E_{C}}{2}\delta N_{g,k}(t)$.

\section{Generation of Colored Noises}\label{App_B}

In this section, we briefly introduce the generation of colored noise fields. The stochastic white fields $\{\eta_{k}(t);k=1,...,N\}$ may be generally written in the form of real white Gaussian process $h(t)$ with $\Braket{h(t)}_{s}=0$ and $\Braket{h(t)h(t')}_{s}=\delta(t-t')$. The colored noise fields can be obtained by digitally filtering $h(t)$~\cite{RJAV:Zhivomirov2018}. The specific algorithm for generating the $\frac{1}{f^{\alpha}}$ noise has four basic steps: (i) The continuous signal $h(t)$ is discretized into a sequence with the sampling period $\frac{1}{f_{0}}$, i.e., $\{\tilde{h}_{n}=h(\frac{n-1}{f_{0}});n=\{1,...,2n_{\textrm{max}}=t_{\textrm{max}}f_{0}\}$; (ii) This finite discrete sequence is then converted into a same-length complex-valued sequence via the discrete Fourier transform,
\begin{equation}
\tilde{H}_{k}=\frac{1}{2n_{\textrm{max}}}\sum_{n=1}^{2n_{\textrm{max}}}\tilde{h}_{n}e^{-i\pi\frac{(n-1)(k-1)}{n_{\textrm{max}}}}.
\end{equation}
The sequence $\{\tilde{H}_{k};k=1,...,2n_{max}\}$ owns the Hermitian symmetry, $\tilde{H}^{\ast}_{k}=\tilde{H}_{2n_{\textrm{max}}-k}$. The linear curve fitting of $\{\tilde{H}_{k};k=1,...,n_{\textrm{max}}\}$ leads to $\tilde{H}_{k}=a_{1}k+a_{2}$ with $a_{1}=0$ and the constant $a_{2}$, corresponding the white-noise spectrum; (iii) Each component in the sequence $\{\tilde{H}_{k};k=1,...,n_{\textrm{max}}\}$ multiples a $k$-dependent factor $(\frac{n_{max}}{k})^{\frac{\alpha}{2}}$, resulting in a new sequence $\{\tilde{H}'_{k}=(\frac{n_{\textrm{max}}}{k})^{\frac{\alpha}{2}}\tilde{H}_{k};k=1,...,n_{\textrm{max}}\}$ and further $\{\tilde{H}'_{k}=\{\tilde{H}'^{\ast}_{2n_{\textrm{max}}-k};k=n_{\textrm{max}}+1,...,2n_{\textrm{max}}\}$; and (iv) Finally, the $\frac{1}{f^{\alpha}}$ noise sequence $\tilde{h}'_{n}$ in the time domain is given by the inverse discrete Fourier transform,
\begin{equation}
\tilde{h}'_{n}=\sum_{k=1}^{2n_{\textrm{max}}}\tilde{H}'_{k}e^{i\pi\frac{(n-1)(k-1)}{n_{\textrm{max}}}}.
\end{equation}
It is seen that $\tilde{H}'_{k}=\tilde{H}_{k}$ for $k=n_{\textrm{max}}$, i.e., the white and colored noise sequences have the same spectral density at the Nyquist frequency $\frac{f_{0}}{2}$.

\section{Metastability}\label{App_C}

\begin{figure}
\includegraphics[width=8.5cm]{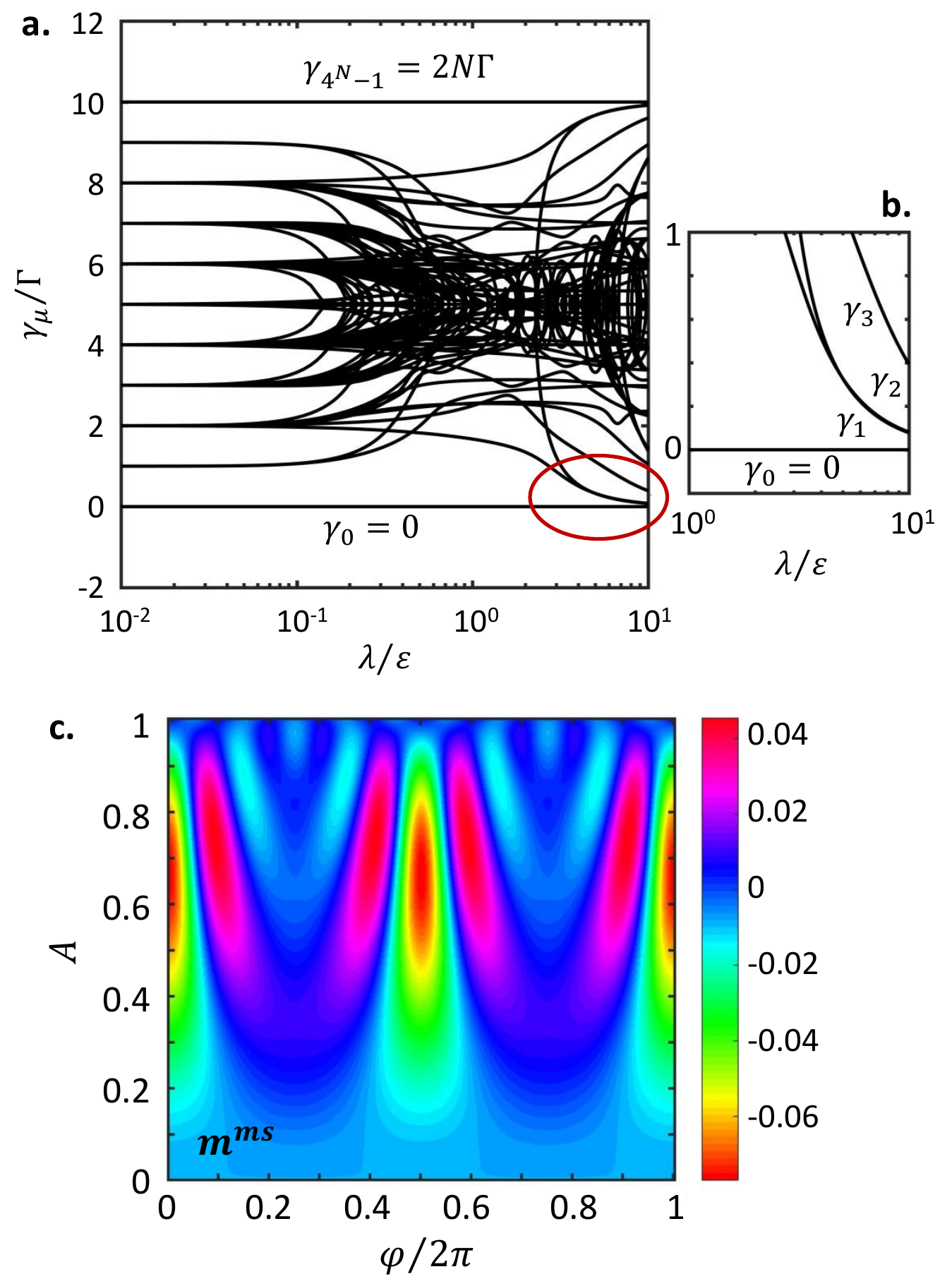}\\
\caption{(Color online) (a) Decay rates $\gamma_{\mu}$ of different relaxation modes vs. the spin-spin interaction $\lambda$. The qubit number is set at $N=5$ because of the limit of matrix dimension that is numerically diagonalizable. The detail in the circle is displayed in (b). Both $\gamma_{0}$ and $\gamma_{4^{N}-1}$ are independent on $\lambda$. The first several lowest $\gamma_{0,1,2,...}$ in the strong interaction regime determine the system's stationary state and metastability. (c) Metastable value $m^{ms}$ as a function of initial state $\prod_{k}\otimes(\sqrt{1-A^{2}}\Ket{\uparrow}+Ae^{i\varphi}\Ket{\downarrow})_{k}$ with $\lambda/\epsilon=10$. All spins are initially prepared in the same state and $\rho(0)=\prod_{k}\otimes((1-A^{2})\Ket{\uparrow}\Bra{\uparrow}+A\sqrt{1-A^{2}}e^{i\varphi}\Ket{\uparrow}\Bra{\uparrow}+A\sqrt{1-A^{2}}e^{-i\varphi}\Ket{\uparrow}\Bra{\downarrow}+A^{2}\Ket{\downarrow}\Bra{\downarrow})_{k}$. The amplitude $A$ varies between 0 and 1 while the phase changes within the range of $[0,2\pi)$.}\label{Fig6}
\end{figure}

We are now in the position to explain the metastability of the dissipative many-body system, for which we solve the Markovian master equation~(\ref{MarkovianMasterEq}) based on the exact diagonalization method. In the product-state basis $\Ket{u}=\prod_{k}\otimes\Ket{v}_{k}$ with $(u=0,...,2^{N}-1)$ and ($v=\uparrow$ and $\downarrow$), Eq.~(\ref{MarkovianMasterEq}) is re-expressed as
\begin{equation}\label{LinearEq1}
\frac{d}{dt}\Bra{u_{1}}\rho\Ket{u_{2}}=\sum_{u_{3},u_{4}}{\cal{L}}_{(u_{1},u_{2}),(u_{3},u_{4})}\Bra{u_{3}}\rho\Ket{u_{4}},
\end{equation}
where we have defined
\begin{eqnarray}
\nonumber {\cal{L}}_{(u_{1},u_{2}),(u_{3},u_{4})}&=&\Bra{u_{1}}\left[-i\frac{H_{s}}{\hbar}-\frac{\Gamma}{2}\sum_{k}(\sigma^{x}_{k})^{2}\right]\Ket{u_{3}}\delta_{u_{2},u_{4}}\\
\nonumber&&+\Bra{u_{4}}\left[i\frac{H_{s}}{\hbar}-\frac{\Gamma}{2}\sum_{k}(\sigma^{x}_{k})^{2}\right]\Ket{u_{2}}\delta_{u_{1},u_{3}}\\
&&+\Gamma\sum_{k}\Bra{u_{1}}\sigma^{x}_{k}\Ket{u_{3}}\Bra{u_{4}}\sigma^{x}_{k}\Ket{u_{2}}.
\end{eqnarray}
Further, we replace the pairs $(u_{1},u_{2})$ with a number sequence $\mu=0,1,...,4^{N}-1$ and Eq.~(\ref{LinearEq1}) is rewritten in the matrix form
\begin{equation}\label{LinearEq2}
\frac{d}{dt}{\bf{R}}=-{\bf{M}}{\bf{R}}.
\end{equation}
The elements of the column vector $\bf{R}$ and $4^{N}$-by-$4^{N}$ matrix $\bf{M}$ are given by $R_{\mu}=\Bra{u_{1}}\rho\Ket{u_{2}}$ and $M_{\mu_{1},\mu_{2}}=-{\cal{L}}_{(u_{1},u_{2}),(u_{3},u_{4})}$. Solving Eq~(\ref{LinearEq2}) leads to
\begin{equation}
{\bf{R}}(t)={\bf{D}}^{-1}e^{-{\bf{E}}t}{\bf{D}}{\bf{R}}(0).
\end{equation}
The diagonal matrix ${\bf{E}}={\bf{D}}{\bf{M}}{\bf{D}}^{-1}$ with $E_{\mu_{1},\mu_{2}}=(\gamma_{\mu_{1}}-i\beta_{\mu_{1}})\delta_{\mu_{1},\mu_{2}}$ lists the eigenvalues of ${\bf{M}}$. The column vector ${\bf{R}}(0)$ denotes to the density matrix $\rho$ at $t=0$. Finally, one arrives at
\begin{equation}\label{LinearEq3}
\Bra{u_{1}}\rho(t)\Ket{u_{2}}=\sum_{u_{3},u_{4}}({\bf{D}}^{-1}e^{-{\bf{E}}t}{\bf{D}})_{(u_{1},u_{2}),(u_{3},u_{4})}\Bra{u_{3}}\rho(0)\Ket{u_{4}}.
\end{equation}
We term $\gamma_{\mu}-i\beta_{\mu}$ the $\mu$-th relaxation mode. The real parts $\gamma_{\mu}$, which are sorted in descending order $\gamma_{\mu}\leq\gamma_{\mu+1}$, denote the decay rates of different modes, while the imaginary parts are related to the interaction-induced level shifts. 

The lowest $\gamma_{0}$ stays zero (Fig.~\ref{Fig6}a), giving the true stationary state of $\rho$, i.e., $\rho^{ss}=\rho(t\rightarrow\infty)$ and $m^{ss}=N^{-1}\sum_{k}\Tr(\sigma_{k}^{z}\rho^{ss})$. The maximum $\gamma_{4^{N}-1}$ is independent on the coupling strength $\lambda$ and equal to $2N\Gamma$. In the weak-coupling limit $\lambda/\epsilon\sim0$, $\gamma_{\mu}$ are divided into $2N$ groups with the equal interval $\Gamma$. As $\lambda/\epsilon$ is increased, $\gamma_{0<\mu<4^{N}-1}$ are strongly modulated, where some modes are pushed towards $\gamma_{0}$ or $\gamma_{4^{N}-1}$. In the strong coupling regime $\lambda/\epsilon\gg1$, the first several decay rates $\gamma_{\mu=1,2,..}$ are close to $\gamma_{0}=0$ (Fig.~\ref{Fig6}b). These modes determine the long-term relaxation dynamics of the system, giving rise to the metastability. In addition, Eq.~(\ref{LinearEq3}) indicates that the metastable behavior of the system depends also on the initial state $\rho(0)$ (Fig.~\ref{Fig6}c).

\end{document}